\documentclass[pre,twocolumn,floatfix,superscriptaddress,showpacs]{revtex4-1}
\usepackage{amsmath,amssymb,latexsym} 
\usepackage{rotating,lipsum}

\usepackage{graphicx}
\usepackage{color}

\begin{document}

\newcommand{\udem}{D\'{e}partement de Physique, Universit\'{e}
de Montr\'{e}al, Montr\'{e}al, Qu\'{e}bec, Canada H3C~3J7}
\newcommand{\nrc}{National Research Council of Canada, Ottawa, Ontario.
 Canada K1A 0R6}

\title
{Ionic structure, Liquid-liquid phase transitions,
X-Ray diffraction, and X-Ray Thomson scattering in
shock compressed liquid Silicon in the 100-200 GPa regime.
}

\author
{
 M.W.C. Dharma-wardana}
\email[Email address:\ ]{chandre.dharma-wardana@nrc-cnrc.gc.ca}
\affiliation{\nrc}\affiliation{\udem}
\author{Dennis D. Klug}
\email[Email address:\ ]{ddklugca@gamil.com}
\affiliation{\nrc}
\author
{Hannah Poole} 
\email[Email address:\ ]{hannah.poole@rochester.edu}
\author{G.Gregori}
\email[Email address:\ ]{gianluca.gregorie@physics.ox.ac.uk} 
\affiliation{ Department of Physics, University of Oxford, Oxford OX1 3PU, United Kingdom}

\begin{abstract} Recent cutting-edge experiments have provided 
{\it in situ} structure characterization and measurements of the
pressure ($P$), density ($\bar{\rho}$) and  temperature ($T$) of shock
 compressed silicon in the 100 GPa range of pressures and up
 to $\sim$10,000K. We present
first-principles calculations in this $P,T,\bar{\rho}$ regime to reveal
 a plethora of novel liquid-liquid phase transitions (LPTs)
 identifiable via discontinuities
 in the pressure and the compressibility. 
 Evidence for the presence of a highly-correlated
liquid (CL) phase, 
as well as a normal-liquid (NL) phase at the LPTs
 is presented by a detailed study
of one LPT. The LPTs make
the interpretation of these experiments more challenging. 
The LPTs preserve the short-ranged ionic structure of the fluid 
by collective adjustments of many distant atoms when subject
to compression and heating, with minimal change in the ion-ion pair-distribution
 functions, and in transport properties such as the electrical
and thermal conductivities $\sigma$ and $\kappa$. We match the experimental
X-Ray Thomson scattering  and  X-ray diffraction data theoretically,
and  provide pressure isotherms, ionization data and compressibilities
 that support the above picture of liquid silicon as a highly complex
 LPT-driven ``glassy'' metallic liquid.
These novel results are relevant to materials research, 
studies of planetary interiors, high-energy-density
physics, and in laser-fusion studies. 
\end{abstract}

%
\maketitle
\section{Introduction}
Laboratory-based study of extreme states of matter, as found in
planetary interiors, or matter produced in high-energy
laser-matter interactions etc., depends crucially on the
development of simultaneous-measurement facilities coupled
to advanced instruments, e.g., such as the Linac Coherent Light Source and the
the OMEGA-EP laser facility~\cite{Poole24}. The characterization of these
 states of matter, usually named ``warm-dense matter'' 
(WDM)~\cite{Ng05,McBride-Si-19},
by simultaneously measuring their pressure ($P$), temperature ($T$),
and density ($\bar{\rho}$), as well as the ionic and electronic structure
is a necessary step for testing the applicability of available
theories of WDM. These theories, crucial even to nuclear-stockpile
stewardship programs, are the only means of
going beyond the reach of laboratory-based studies which are limited
by the energy scales of terrestrial facilities.

Si, together with C, P, As etc., are light elements that are
  insulators or  semiconductors that become metallic liquids when molten,
with an increase in density.
 They manifest transient covalent binding even
after melting~\cite{Si-Aptekar79,Stich89, DWP-carb90,
 OkadaSi12,CPP-carb18}. Furthermore, Si is of fundamental importance
in the physics of planetary interiors, and in many areas of
materials science, device fabrication and shock physics~\cite{sxhu2017}.
The electrical and thermal conductivities of $l$-silicon are also
of interest in the context of planetary magnetism and thermal balance. 

It's fundamental physics is also intriguing and challenging due
 to the presence of  many liquid-liquid phase transitions
 (LPTs)~\cite{cdwSi20,SastryAngel03}.
 The LPTs are also known as plasma-phase transitions. The LPTs in liquid-Si 
include a well-studied transition between a high-density branch of the liquid
and the low-density branch  at a density of $\sim$2.27-2.5 g/cm$^3$
at $\sim$1200-1800K, near its melting point. A highly-correlated
 phase appears near the LPT and  persists even into the supercooled
regime (1200K)~\cite{SastryAngel03,Ashwin04,Morishita04,McMillan05,
Morishita06,Beaucage05, Daisen07SiPhDia,GaneshSiLPPT-09,Baye10,
Vasisht11,Remsing17,Remsing18}. We refer to this LPT as LPT(2.5$|$1200K)
in the foregoing.
The persistence of these LPTs even to temperatures as high as 1 eV
(11,604K) was predicted in Ref.~\cite{cdwSi20} but experimental 
studies of such WDM states have been lacking.

The recent study of shock-compressed liquid silicon ($l$-Si) by Poole
et al~\cite{Poole24} in the 100 GPa regime has provided simultaneously
 measured data for X-Ray Thomson scattering (XRTS) and X-Ray diffraction
 (XRD), as well as pressure and temperature determinations. They conclude
that the WDM-Si created was  at a density of 4.56$\pm0.07$ g/cm$^3$, with
a pressure of 106$\pm$6 GPa. However, these estimates need to be re-examined
using a first-principles atomic physics model.

A first-principles  approach to the analysis of such data
 is to use $N$-atom DFT calculations, 
 coupled to molecular dynamics (MD) simulations~\cite{ABINIT,VASP},
 a method referred to here  as `quantum molecular dynamics' (QMD).
 It is also sometimes referred to as MD-DFT in the literature.
  Poole et al~\cite{Poole24} did not use QMD, pointing out that
 ``such methodologies~\cite{Schorner23,Witte17}
 are computationally expensive and introduce varying degrees of
 complexity''. Their study was conceived more as a proof-of-principle
application of a multi-messenger diagnostic approach.
As such, to maintain fast computations within a Markov-Chain Monte Carlo
analysis, they opted for a simple pair-potential
model~\cite{Varshni90} that uses a non-linear  Hulth\'{e}n (NLH) potential. 
They analyzed the XRD and XRTS data, with the pressure
estimated from  an equation of state (EOS)
model due to Vorberger et al~\cite{Vorberger13}, and 
Ebeling et al~\cite{Ebeling20}. They concluded using only  velocity interferometry
measurements (VISAR) that the $l$-Si 
had a density of $4.43\pm$ 0.8 g/cm$^3$, revised using XRD data to
 $\sim$4.6 g/cm$^3$, with $T$ =5300K, and a charge state with
$\bar{Z}=1.3$, i.e., each `average' Si atom has
given up 1.3 electrons to become Si$^{1.3+}$ ions.   

In QMD,
many silicon atoms, e.g., $N=64-216$ or more are used to capture
 many-ion (Si-Si) interactions, while an electron XC-functional
 reduces the many-electron problem to an effective ``non-interacting'', i.e.,
 ``single-electron''
model as in a Lorentz plasma~\cite{LLvol10} where only electron-ion
interactions need to be dealt with.
 In the same way, the $N$-ion
problem can be reduced to an effective ``one-ion'' problem by using an
 XC-functional suitable for ion-ion interactions~\cite{DWP82}. The XC-functionals
 used in the NPA are discussed in more detail in the appendix.
 It is this effective ``one-ion'' DFT approach, known as the neutral-pseudo-atom
(NPA) approach~\cite{DWP82,Pe-Be93,eos95,DagensNaLi72} that we use in our calculations.
  A recent account of the NPA method may be found in Ref.~\cite{cdwSi20}.
It is similar to an average-atom (AA) model and hence computationally
 very economical, unlike QMD. We also ascertain, using limited
QMD calculations, that the structure data obtained from the NPA and from
the QMD are in good agreement.

Using the NPA technique we calculate the structure factors, pair-distribution
functions (PDFs), pressure isotherms, comprehensibility electrical and
 thermal conductivities  $\sigma, \kappa$
for $l$-Si for densities within
 4.0 g/cm$^3$-5.0 g/cm$^3$. The main peaks of the 
calculated  structure factors $S_{ii}(k)$ are found to be
nearly self-similar and located very nearly at $k_1=1.5k_F$,
  where $k_F$ is the Fermi wave vector. As discussed in the
Appendix, the main peak of $S(k)$ gets  positioned
to take advantage of the zero of the electron-ion pseudopotential $U_{ei}(k)$.
The interaction potentials are very long-ranged due to Friedel tails. Hence
all charge distributions are calculated to about ten Wigner-Seitz radii or more.
When the fluid is compressed or heated, small collective displacements of many
 distant ions occur to  preserve the  short-ranged ordering,  
by inducing many LPTs and discontinuities in the
 compressibility and the pressure. 

These LPTs have only a minimal impact 
on the conductivites $\sigma,\kappa$. This arises since
 the  main peak of $S_{ii}(k)$, and the
window of interaction of the electron-ion pseudopotential around 2$k_F$,
 remain largely unaffected in the range of small $t=T/E_F$, where
$E_F$ is the Fermi energy. Thus, for $\bar{\rho}$ =4.6 g/cm$^3, T=$ 1 eV,
$t=0.051$. 

It is hard to detect LPTs using QMD simulations (unless they directly
affect short-ranged ordering) because, even in a
 simulation using 216 atoms; the linear scale is about 6 Wigner-Seitz radii.
 The central atom has no more than two to three neighbours on each
 side. The interionic potentials generated in such QMD simulations
 are thus very short ranged, compared to those from NPA
calculations, as was demonstrated in Ref.~\cite{DW-Yuk22},
 for the pair-potentials for carbon extracted from 
QMD calculations of Whitley et al~\cite{Whitley15}. Machine-learnt
potentials~\cite{ercolessi1994interatomic,Brommer2015Potfit,Stanek21}
 using multi-ion models (e.g., extensions of the
Stillinger-Weber model or effective-medium models), bond-order potentials etc.,
 have not been successful in their applications
 to these WDM systems~\cite{kraus13,DW-Yuk22}. 
 A similar problem arises in the study of
Martensitic transitions in iron~\cite{KadauFe2002} where small collective
modifications in positions of many `far-away' ions interacting via
 long-ranged potentials occur. In contrast, if significant changes
in short-range order occur at an LPT, then QMD calculations with
even 64 atoms using the PBE functional may be sufficient. However
simulations with 216 atoms and the SCAN functional are recommended
 in Remsing  et  al.~\cite{Remsing17}, for instance, to obtain quantitative
 predictions of the melt line of silicon.

Besides the LPTs that preserve the short-ranged order,
there are also correlated liquid phases that appear
 between the high-density branch and the low-density
 branch that form an LPT. Here even  supercooled
 persistent states, identified as  somewhat solid-like 
correlated liquid (CL) phases have been revealed in
dynamic simulations~\cite{Remsing18}, as in the LPT at 2.5 g/cm$^3$ near
the melting point of Si.
  
In this study, directed more to a reinterpretation of the XRTS
and XRD data, a full analysis of the LPTs will not be
attempted. Instead, we will study just one of them in
some detail, to report the observation of a correlated
liquid state and to bring out the similarities to the well-studied
LPT(2.5$|$1200K) of normal-density Si near its melting point~\cite{cdwSi20}.

The presence of LPTs in $l$-Si renders it highly viscous, prone to hysteresis,
 and ``glassy'' to shock propagation, slowing down shock speeds. This may lead
  to lower estimates of pressure and temperature for shocked media if the LPTs
  are ignored. Our calculations simulate the XRD and XRTS data 
of Poole et al~\cite{Poole24} and show that
their estimate of $T$=0.4567 eV (5300K) and a pressure of 100 GPa,
 with $\bar{Z}$=1.3 are indeed lower estimates compared to ours.
We find  $T\sim$  1-1.2 eV, $P\sim$ 150 GPa, and $\bar{Z}$=4.

We present first-principles pressure ($P$), compressibility ($\xi$), 
transport ($\sigma,\kappa$), and ionic structure data,
as well as XRTS and XRD intensity spectra. The notation for ion-ion data will be 
simplified in the foregoing  to, e.g., $S(k)=S_{ii}(k), g(r)=g_{ii}(k)$ when no
 confusion is likely to arise.  The theoretical results  show
good agreement with the available experimental data and
 QMD simulations of structure data.
Additional details of these calculations are given in
the Appendix.

\begin{figure}[t]                    
\includegraphics[width=0.96\columnwidth]{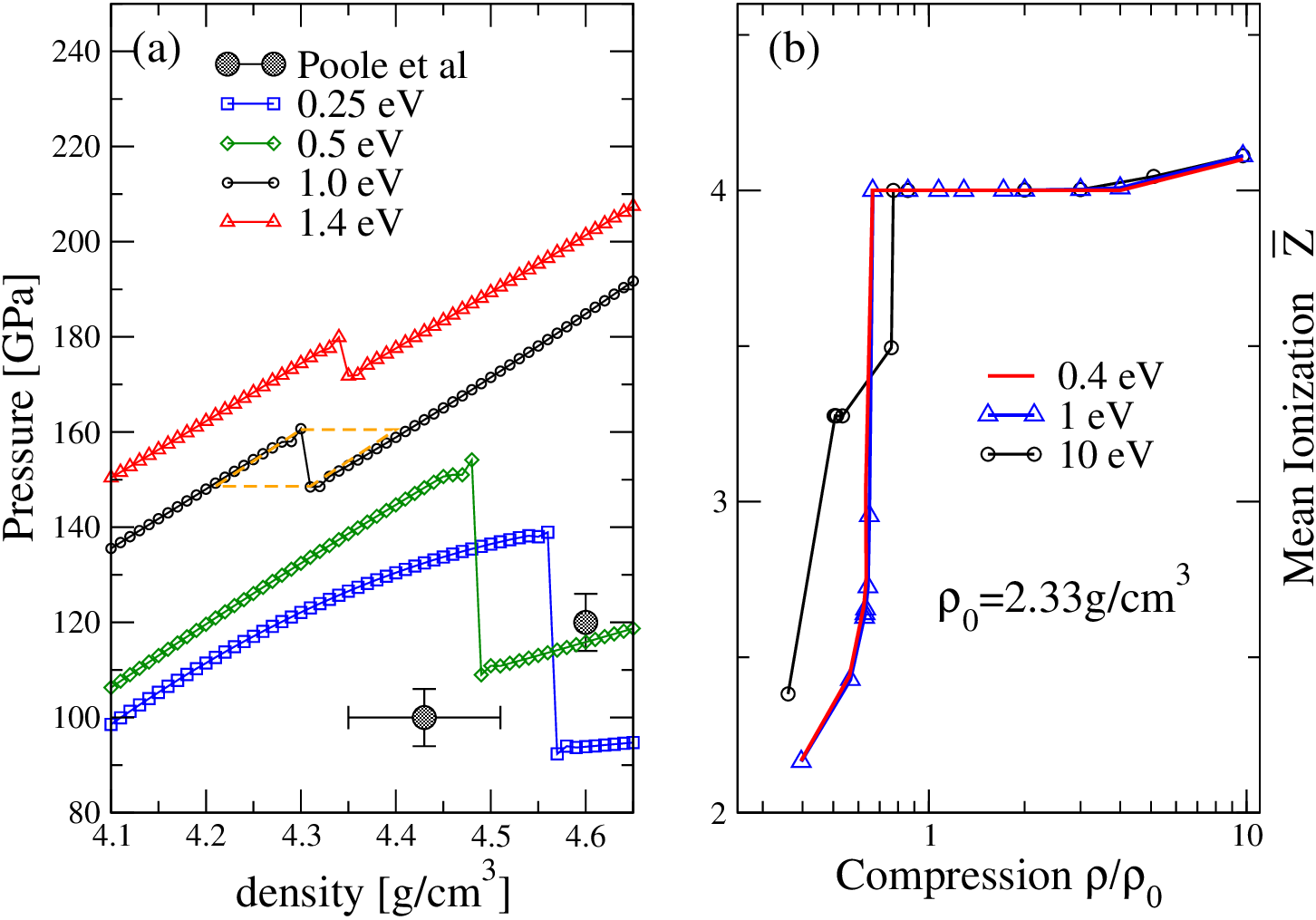}
 \caption{\label{rho-p.fig}(online color) EOS data implying large
hysteresis in compression/expansion, and in heating/cooling processes
 in $l$-Si. (a) Pressure isotherms in
 the density range 4.1-4.7 g/cm$^3$  display many discontinuities due to
liquid-liquid phase transitions~\cite{cdwSi20}
 The data point derived from the experiment, Ref.~\cite{Poole24},
using their atomic model and only VISAR measurements  (4.43 g/cm$^3$),
and the data point for final density of 4.6 g/cm$^3$ based on
 VISAR, XRD and XRTS are also shown.
(b)The mean ionization $\bar{Z}$ as a function of compression
 $\bar{\rho}/\rho_0$, where $\rho_0$=2.33 g/cm$^3$.}
\end{figure}

\begin{figure}[t]                    
\includegraphics[width=0.96\columnwidth]{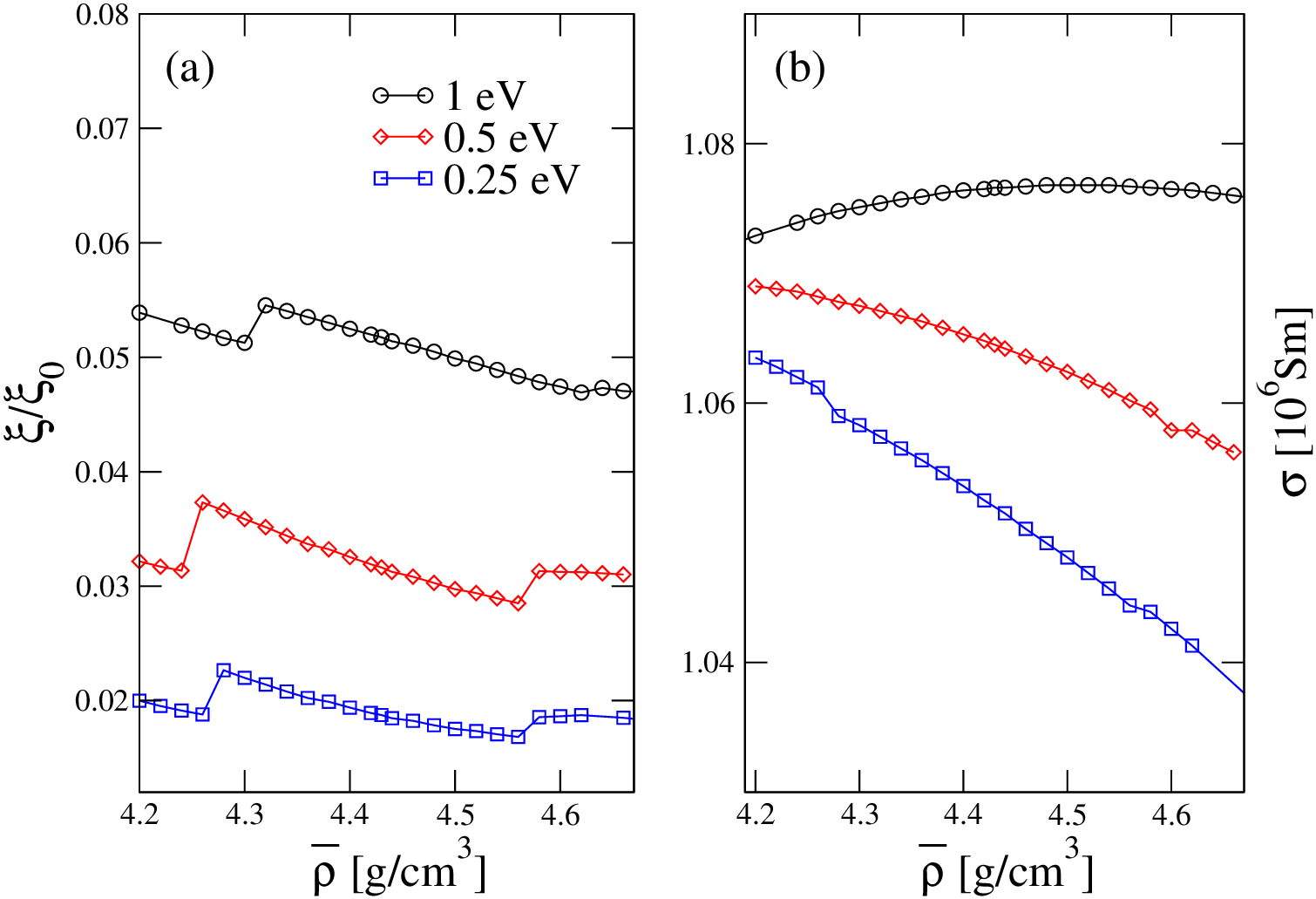}
 \caption{\label{s0.fig}(online color) (a) Discontinuities in $S(0)=\xi/\xi_0$,
the compressibility ratio, independently confirms the LPTs.
(b) The electrical conductivity $\sigma$ is only very weakly affected by the LPTs.
At $T=025$ eV, $\sigma$ changes by less than 3\% when the density changes by about 25\%.
At $T= 1$ eV the change is only a fraction of a percent (c.f., Appendix). }
\end{figure}

\section{ Structure data and EOS} 
We use Hartree atomic units
$\hbar=|e|=m_e=1$ in a standard notation. The electron-sphere radius
$r_{s}=[3/(4\pi\bar{n})]^{1/3}$, the ion Wigner-Seitz radius 
$r_{ws}=\bar{Z}^{1/3}r_s$, and the Fermi wavevector $k_F=(2\pi^2\bar{n})^{1/3}$
are used in our equations and graphics.

As discussed in more detail in the Appendix,
the  NPA provides electron-ion pseudopotentials $U_{ei}(k)$, pair potentials
$V_{ii}(r)$, pair-distribution functions (PDFs) $g(r)$,
 the structure factor $S(k)$. It provides
 all  electronic-structure data  necessary for the calculation of the
 total Helmholtz free energy of the
 system as well as all other properties including XRTS data and transport
 properties, from first principles. 

The isothermal compressibility $\xi$ of the fluid
 is directly available from  the $k\to 0$ limit of $S(k)$, viz., $S(0)=\xi/\xi_0$,
where $\xi_0=1/(\bar{\rho}T)$ is the ideal-gas compressibility.  Discontinuities
 in the pressure and the $S(0)$ reveal the presence of a plethora of
 liquid-liquid phase transitions. 
Unlike in QMD where the small-$k$ limit of $S(k)$ is not
accessible due to the finite size of the simulation cell, the
large radius of the correlation sphere used in
the NPA enables an accurate evaluation of the $S(0)$ limit.

Fig.~\ref{rho-p.fig} provide EOS data for $l$-Si indicating abrupt breaks
due to LPTs similar to the well-know transition near $\sim$2.2-2.5 g/cm$^3$ at
the melting line~\cite{cdwSi20, Remsing17}. The figure also shows how
the experimental data point (see Table III~\cite{Poole24}) evolves when only VISAR
data are used in the analysis, and when a more complete set of diagnostics
(XRD, XRTS) is used in the analysis.

In Fig.~\ref{s0.fig}(a) we display
the compressibility ratio $\xi/\xi_0$ of $l$-Si, confirming the LPTs observed in
the $P$-isotherms. Usually, phase transitions also imply abrupt changes in
the electrical and thermal conductivities if changes occur at the Fermi surface. 
What is noteworthy here is that while the density changes from $\sim$ 4 g/cm$^3$
 to $\sim 5$ g/cm$^3$, i.e., a change of 25\%, the electrical conductivity
  changes less than
3\% at $T=0.25$ eV, while at 1 eV, the change is less than 0.3\%. The same is
true for the thermal conductivity given in table~\ref{xisigkap.tab} of the appendix.
Furthermore, as seen in Fig.~\ref{s0.fig}(b),
the LPTs have very little impact on these transport properties. This is well
accounted by our theoretical results which show that (i) the mean ionization
$\bar{Z}$ remains unchanged across these LPTs; (ii) any significant variation
 in the position of the first peak of $S(k)$ is suppressed as it remains tied to
  a zero of the form factor of the pseudopotential, thereby reducing
 the electron-ion interaction energy at the Fermi surface. (iii) Although
a $g(r)$ of a highly-correlated  liquid (CL) is detected near LPTs through
simulations, it seems to have no impact on the conductivity~\cite{cdwSi20}.

Our analysis of the XRTS and XRD data
favour a density of $\sim 4.3$ to 4.4 g/cm$^3$,T=1 to 1.2 eV,
 at a pressure of $\sim 150$ GPa.
This higher $P,T$ compared to the estimate of Ref.~\cite{Poole24}
is consistent with an extremely glassy material, caused by the LPTs. Furthermore,
panel (b) of Fig.~\ref{rho-p.fig} show that a value of $\bar{Z}=1.3$ as in
Ref.~\cite{Poole24} is
not attained even at the lowest density of 0.93 g/cm$^3$, compression $\sim 0.4$
 for temperatures of 0.4 eV to 10 eV. We have confirmed this even at 0.25 eV,
though not displayed in Fig.~\ref{rho-p.fig}. Furthermore, it was shown~\cite{cdwSi20}
 that the Ganesh-Widom pressure data~\cite{GaneshSiLPPT-09} along the melt line of
 $l$-Si at 2.33 g/cm$^3$ is recovered in NPA calculations
with $\bar{Z}=4$ as the mean ionization. We also find that the NHL potential is unable to
generate the known structure data (e.g., $g(r)$) for $l$-Si at, e.g., 2.5 g/cm$^3$
and $T=1800$K, at the $Si$-melt line.

\begin{figure}[t]                    
\includegraphics[width=0.96\columnwidth]{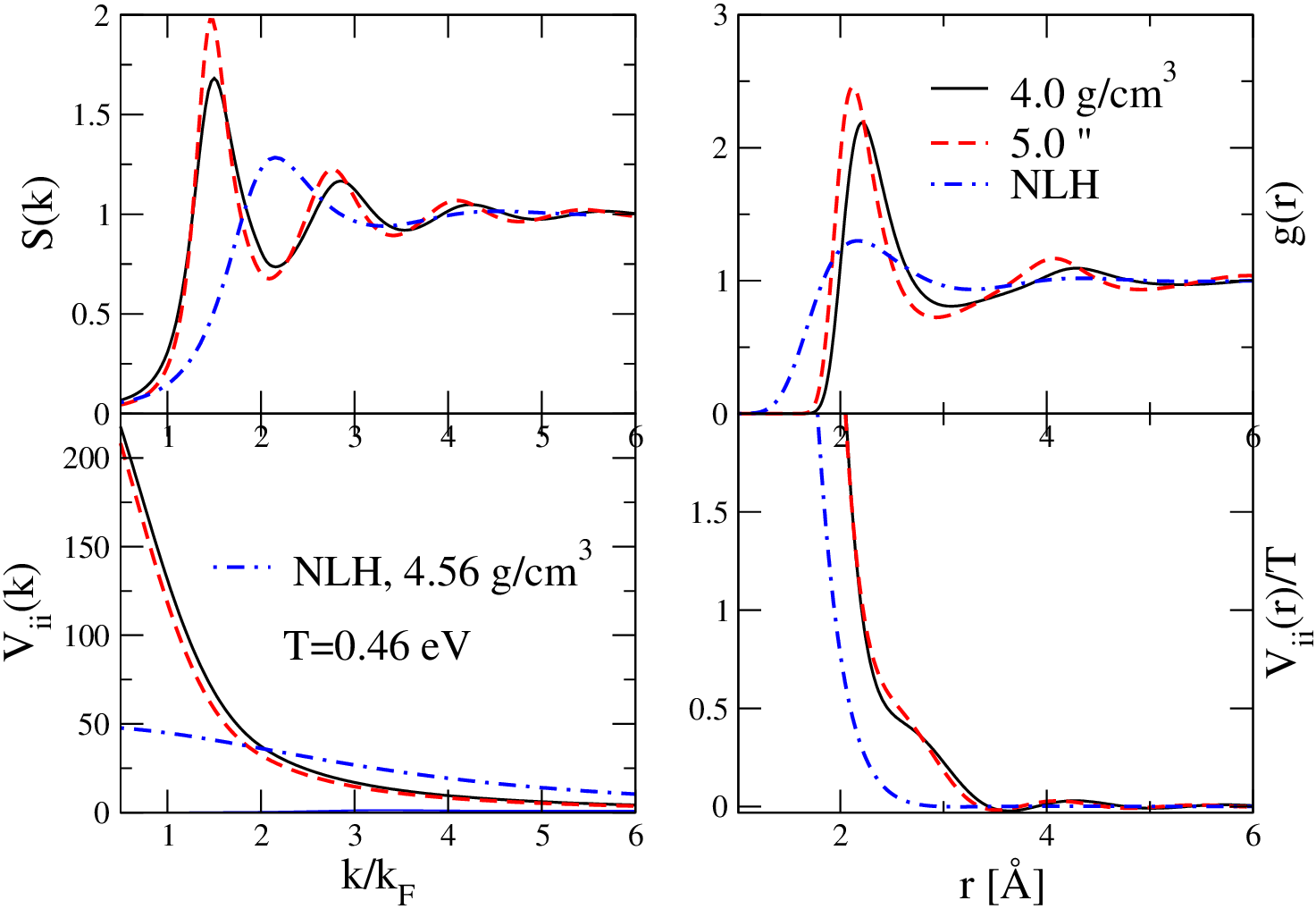}
 \caption{\label{skgrvr.fig}
(online color) Panel (a) The structure factor for
$l$-Si at 4.0 g/cm$^3$, and at 5.0 g/cm$^3$ at T=0.4567 eV obtained from the
NPA approach, compared with the $S(k)$ used in Ref. ~\cite{Poole24} derived from
the NLH potential at 4.56 g/cm$^3$ at T=0.46 eV. The $k$-axis is
scaled with the appropriate $k_F$. (b) Comparison of the PDFs $g(r)$.
(c) Comparison of the pair-potentials in $k$-space. (d) The real-space
potentials $V_{ii}(r)$.}
\end{figure}

The comparison of the potentials and structure data shown in
 Fig.~\ref{skgrvr.fig} shows that there is a large difference between the
NPA results and the NLH model. The latter is closely 
similar to the Yukawa model and hence this lack of agreement with a
first-principles atomic model is not unexpected.

  Furthermore, we show below
 that the XRTS ad XRD profiles are in much better agreement with the NPA
calculation. The $S(k)$ and $g(r)$ calculated
for various WDM conditions using the NPA  method has been shown to agree very
closely with those obtained by QMD simulations in all instances
 where comparisons are available, and more specifically for
 $l$-Si~\cite{cdwSi20,cdw-Utah12} as well. As the current study is concerned with
compressed Si in the regime of density $\bar{\rho}$ 4.3-4.6 g/cm$^3$ and $T$ in the
range 0.46 eV to 1.2 eV,  specific comparisons of the pair-distribution
functions $g(r)$ obtained using the NPA, and QMD calculations using the
 Vienna Ab-initio Simulation code (VASP) were carried out,
 and examples are given in the following section where we discuss in detail the LPT
at 4.3 g/cm$^3$ and $T$ = 1 eV.

\begin{figure}[t]                    
\includegraphics[width=0.92\columnwidth]{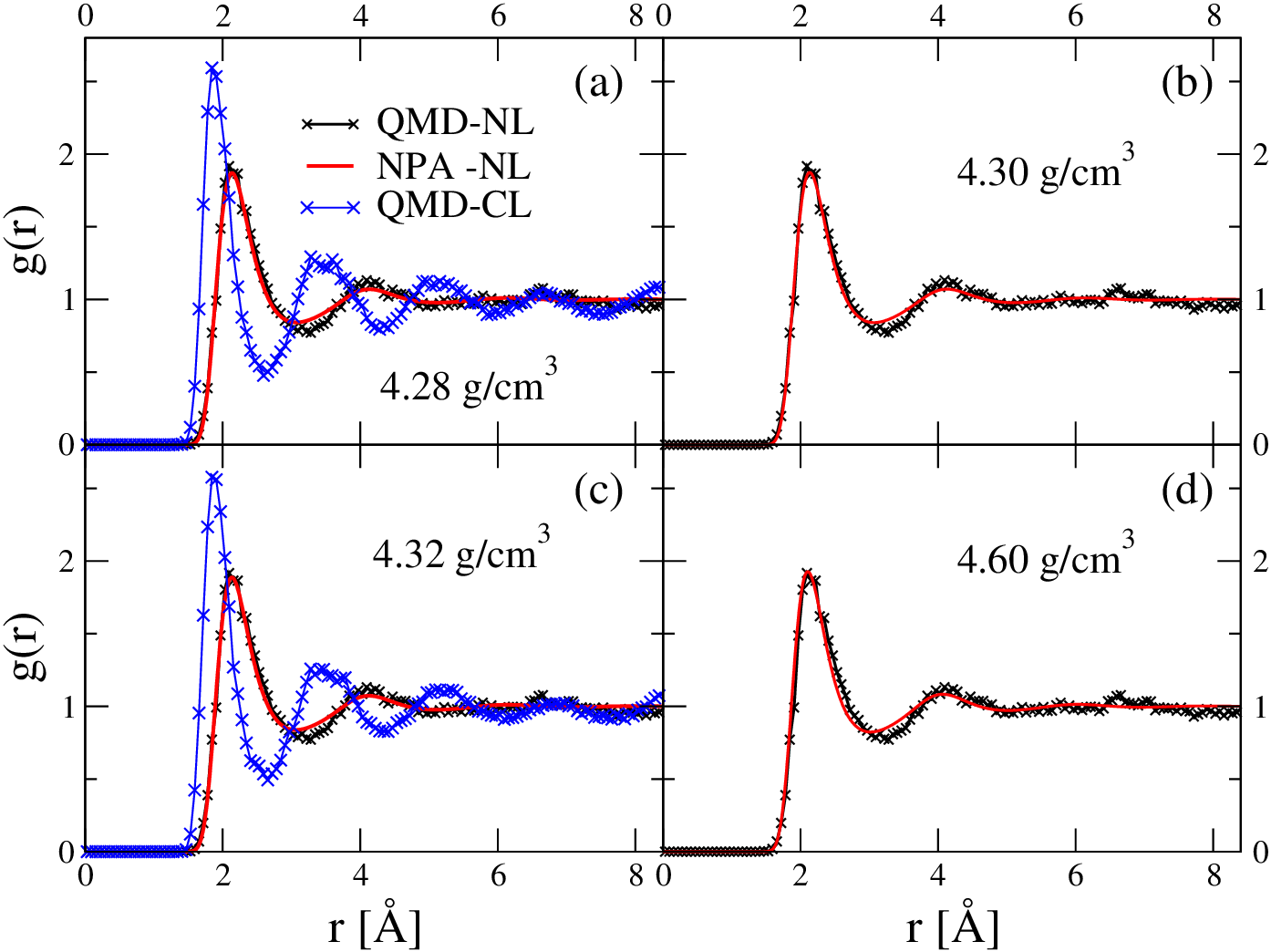}
 \caption{\label{LPT4p3-1ev.fig}(online color) The PDFs $g(r)$ obtained via NPA and from
 QMD-VASP at densities near the LPT at 4.3 g/cm$^3$ and $T$ = 1 eV. 
The NPA and VASP-QMD generated $g(r)$ agree closely for the normal-liquid (NL) phase.
A highly-correlated liquid (CL), with the 1st peak in $g(r)$ closer in ($<2$\AA), and
 is found and displayed in panels (a) and (c). This was not detected at the higher density of 4.6 g/cm$^3$, away from the LPT, where only the normal liquid is found.
}
\end{figure}

\section{The LPT at $4.3\mbox{ g/cm}^3$ and $1\mbox{ eV}$}
\label{LPT4p3.sec}
Fig.~\ref{rho-p.fig} reveals six LPTs on the three isotherms shown
there. A comprehensive study of all the LPTs, as was done
for LPT($2.5|1200$K)~\cite{cdwSi20} making extensive use of QMD,
is computationally costly and beyond
the scope of this work. Instead, we study just the
prominent LPT near $\bar{\rho}$ = 4.3 g/cm$^3$ and $T$
 = 1 eV, viz., LPT($4.3|1$eV) as this is closest to the
 density and temperature revealed by our study of the
 XRTS data discussed in Sec. IV.

\begin{figure}[t]           
\includegraphics[width=0.96\columnwidth]{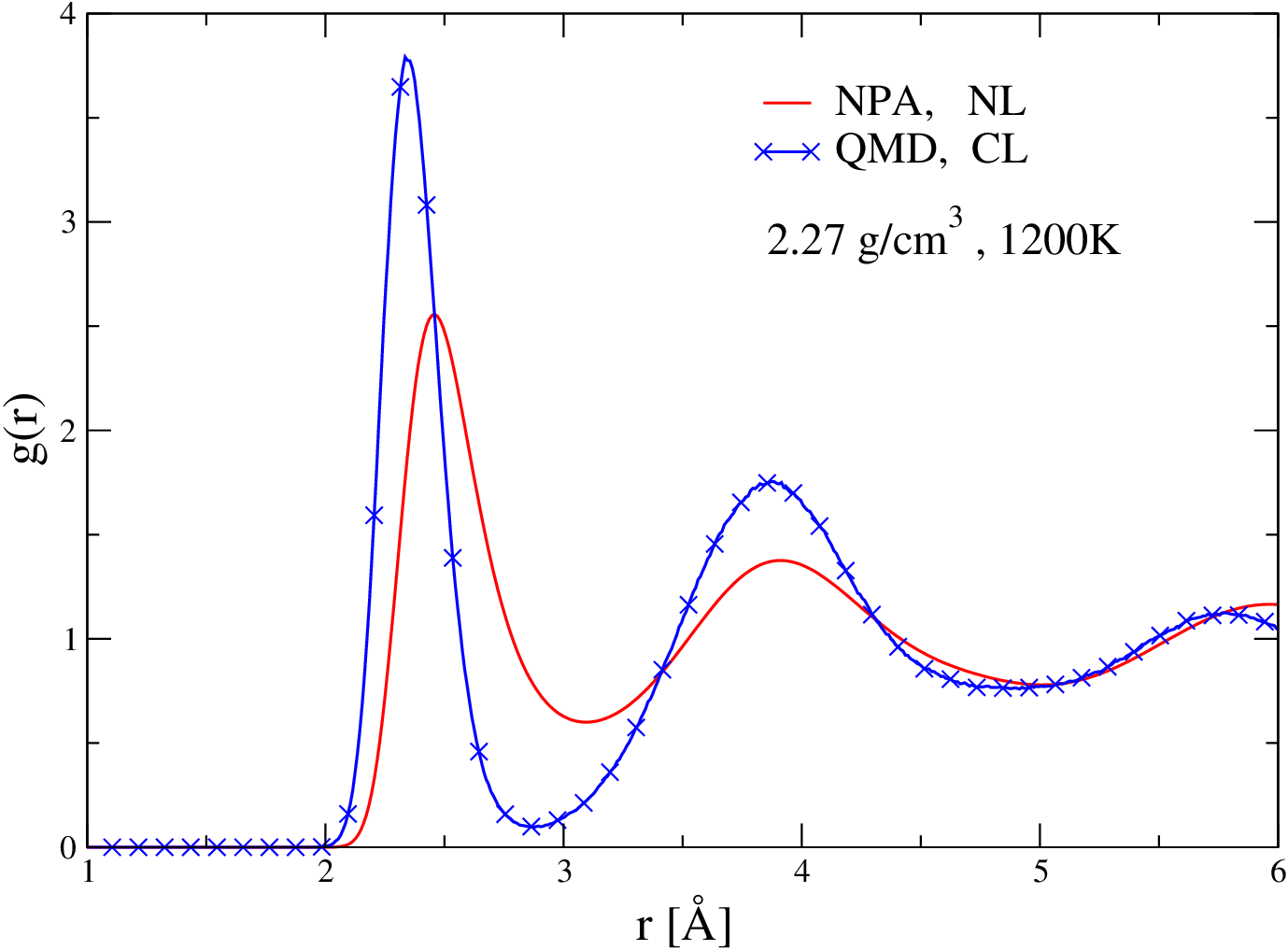}
 \caption{\label{lpt2p7.fig}(online color) The PDFs $g(r)$ obtained via NPA and from 
QMD (216 atoms+SCAN functional~\cite{Remsing17,Remsing18}) at densities near the LPT at 2.27 g/cm$^3$ and $T$ = 1200K.
The NPA and QMD generated $g(r)$ agree closely for the normal-liquid phase.
The highly-correlated liquid (CL) similar to those in Fig.~\ref{LPT4p3-1ev.fig}(c),(d),
 with the 1st peak in $g(r)$ closer-in than the usual
Si-Si bond distance of 2.35 \AA, has been reported in ~\cite{cdwSi20, Remsing18}.
}
\end{figure}

We note that the short-ranged structures of the LPTs
at 1 eV as revealed by the $g(r)$ calculated via the NPA
for the stable liquid phase remain essentially unchanged
across the LPTs and even for the full range of $\bar{\rho}$/gcm$^{-3}$
from 4.28 to 4.6, as seen in Fig. 4. The NPA $g(r)$
and the $g(r)$ calculated via the VASP code agree very
well for the normal form of $l$-Si. Details of VASP calculations
are given in the Appendix. A more correlated liquid phase is
also detected at $\bar{\rho}$/gcm$^{-3}$ = 4.28 and 4.32.
At first sight it is surprising that such a correlated
phase might exist at as high a temperature as 1 eV.
However, $T=1$ eV is only $T/E_F=0.053$ at this compression.

 As with the LPT($2.5|1200$K), the CL-phase
 is difficult to obtain via the NPA which seeks
 the lowest-energy uniform-fluid phase. Whether this correlated
 phase plays a role in
hysteresis loops, and even under supercooling, as was
found for the LPT($2.5|1200$K) using the SCAN functional
and QMD calculations, have not been investigated in this
work. However, we conclude from this limited study that
the LPTs found here are similar to the LPT($2.5|1200$K)
studied in Ref.~\cite{cdwSi20} for $l$-Si. Further discussions
 of the $g(r)$ and $S(k)$ of $l$-Si at the LPTs are given
 in Sec. A 4.

\section{X-Ray diffraction and X-ray Thomson scattering profiles} 
\label{XRTS.sec}
\begin{figure}[t]                    
\includegraphics[width=0.96\columnwidth]{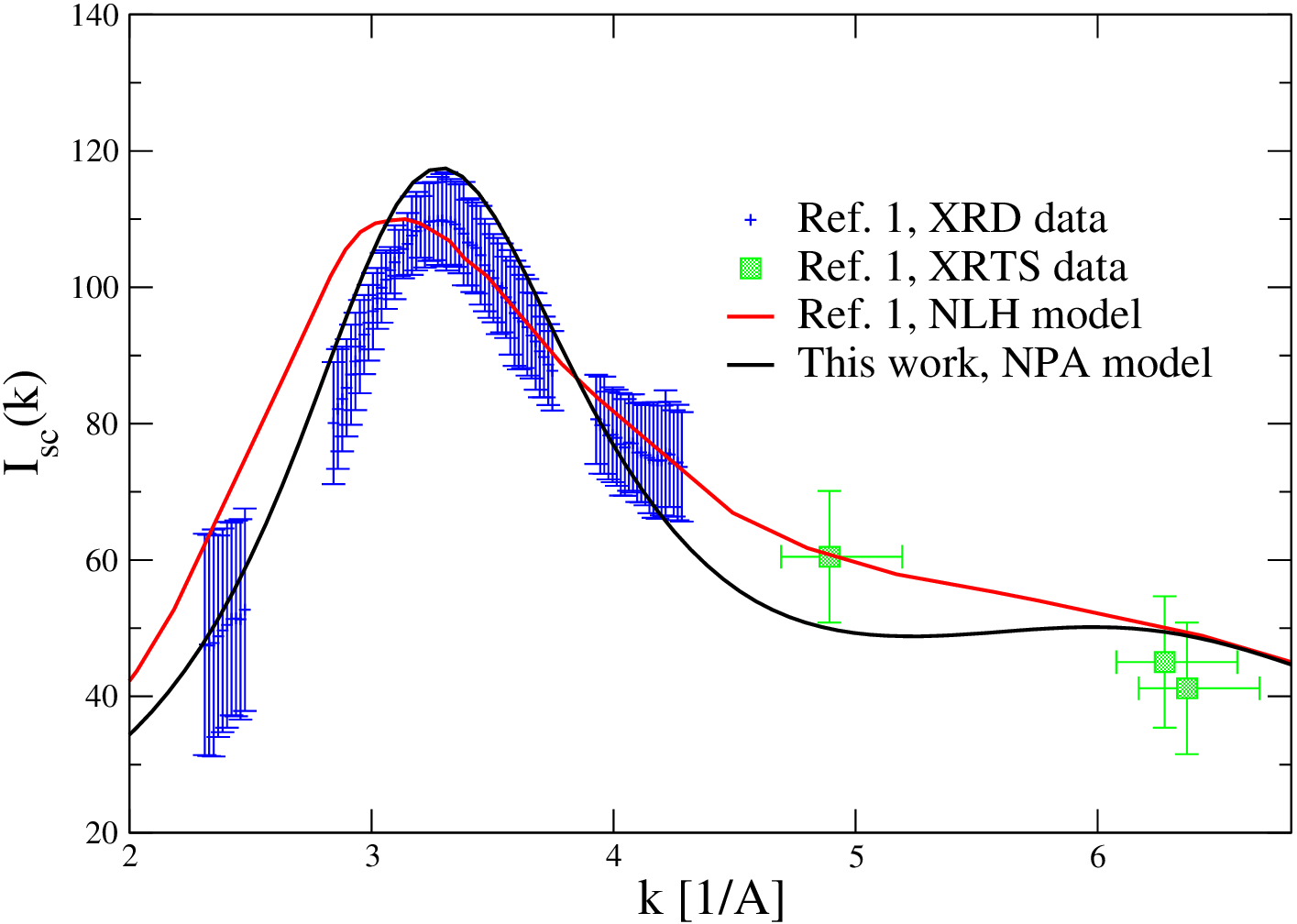}
 \caption{\label{IscPZbar.fig}(online color) The calculated scattered-light intensity
 is compared with the experimental X-ray diffraction (XRD) data and X-ray Thomson
 scattering (XRTS) data. The results for the NLH potential is from Ref.~\cite{Poole24}
with $\bar{\rho}$=4.6 g/cm$^3, \bar{Z}=1.3$, and $T=0.4567$ eV. The NPA calculation is at
$\bar{\rho}$=4.35 g/cm$^3, \bar{Z}=4.0$, and $T$=1.2 eV.}   
\end{figure}

The NPA model has  been successfully used already for analyzing
XRTS and XRD data~\cite{xrt-Harb16}. The Kohn-Sham calculation
 of the NPA provides the Fourier transform
 of the bound, and free
 electron density, viz., $n_b(q)$, $n_f(q)$ respectively.
 The `ion feature' $I_{coh}(k)$ which is proportional to the scattered signal is such that
$I_{coh}(k)=|n_b(k)+n_f(k)|^2S(k)$. The observed signal includes an incoherent
scattering signal $I_{inc}(k)$ as well. The scaled scattering signal
 provided by the experiment of Ref.~\cite{Poole24} and the results
 from the NLH model are shown in Fig.~3(a) of Poole et al.
 We simulate their experimental data (Fig.~\ref{IscPZbar.fig})
using the NPA-calculated ion-feature $W(k)$, to which  $I_{incoh}(k)$
 has been added, following Ref.~\cite{Poole24}. 

The experimental signal $I_{lqd}$ has been scaled
 to match the very high-$k$ limit of a  calculated value of $I_{ic}$
\begin{eqnarray}
I_{lqd}/\gamma &=& I_{coh}\{S(k)-1\}+I_{ic}\\
I_{ic} &=& I_{coh}(k)+I_{incoh}(k)
\end{eqnarray}

\section{Discussion} 
While the NPA-calculated XRD profile matches
the experimental XRD data closely, the intensity is overestimated by $\sim$ 6\%. 
A better fit can be obtained by slightly increasing the temperature of the fluid.
However, given the error bars of the experiment, and the strong likelihood of
hysteresis in this system, further optimization of the fit adds little insight.
In our view, the higher estimate of $P\sim150$ GPa, $T\sim 1.2$ eV, $\bar{\rho}$ 
= 4.35 g/cm$^3$  obtained here, compared to $P\sim100$ GPa, $T=0.46$ eV is
 consistent with $l$-Si being a glassy LPT-ridden liquid rather than a
 simple liquid, as assumed in currently available Si-Hugoniots.  

The existence of transient covalent bonds in liquid metals and plasmas
 such as $l$-carbon, $l$-Si and other low-Z materials has been appreciated
 by many workers since the 1980s.  A description of such bonding states in
 terms of Wannier functions~\cite{Andreoni76} may
 be possible and might be appropriate near the percolation threshold that would occur
 as $\bar{Z}$ decreases with decreasing density, as shown in Fig.~\ref{rho-p.fig}(b).
A Wannier-function description may also be appropriate for modeling the more-ordered
 CL-phase found at LPT(4.3$|$1 eV) and at the LPT(2.5$|$1200K). However, the
normal-liquid  phase is the most persistent form of $l$-Si. The normal
liquid $S(k)$ has been used here, in the interpretation of the X-ray scattering data. Given that
 the WDM-Si studied here is in the strongly metallic regime with $\bar{Z}=4$,
 the Kohn-Sham
 states of the continuum electrons are well represented by
 phase-shifted plane waves. This is attested by the fact that a simple weak $s$-wave
 pseudopotential (constructed {\it in situ} from the displaced Kohn-Sham density) is
 sufficient to describe ion-ion interactions. 

Furthermore, the low ionization of $\bar{Z}\simeq 1.3$ proposed in Ref.~\cite{Poole24}
 is a result of the use of the NLH model within the Markov-Chain Monte-Carlo (MCMC) procedure.
 The latter was developed mainly for analyzing plasmas in the classical regime and
 hence the $\bar{Z}\sim 1.3$ merely played the role of a fitting parameter. 

The result of the NPA calculations are in agreement with the conclusions
 of Ref.~\cite{cdwSi20} that $l$-silicon at $\sim 4-5$ g/cm$^3$ is a good
 metal at 1 eV, with about a third of the conductivity of normal-density Al at the
 same temperature. Snap shots of of ionic positions in QMD simulations
of $l$-Si are suggestive of complex transient bonding similar to that see in $l$-carbon.
 However, the long-time averaged (i.e., static) Si-Si pair distribution function of $l$-Si,
 calculated using the ``one-atom''-DFT approach of the NPA agree well with that obtained from
 QMD calculations, as demonstrated in Sec.~\ref{LPT4p3.sec}, and in Sec. A4 of the
 Appendix detailing QMD simulations.  These results should have implications regarding planetary
 magnetism due to the possibility of electrical currents in molten-Si cores.

In conclusion, the present analysis, using the all-electron DFT approach as
employed in the NPA model, suggests that the WDM sample of $l$-Si created in
 the study by Poole et al~\cite{Poole24} is very likely 
to be at a density of $\sim$ 4.35 g/cm$^3$, at an ionization of four free electrons/ion, 
and in the pressure range 140-160 GPa. The NPA study presents a novel picture of
$l$-silicon as a highly ionized metal showing a plethora of liquid-liquid phase transitions. 
Experimental pressure and temperature estimates  may have significant
uncertainties due to hysteresis effects, posing a further challenge to future experiments
and theory.

\section{Acknowledgments}
The work of GG was partially supported by the United Kingdom EPSRC and First Light
 Fusion under the AMPLIFI Prosperity Partnership -EP/X025373/1. The first two 
 authors (MWCD, DDK) have not received any specific funding for this work.


\appendix
\section{}
$\,$\\
This appendix  addresses the following topics.\\
(i) Brief description of the NPA calculations and the ion XC-functional.\\
(ii) Pseudopotential, Structure data and the first peak of $S(k)$.\\ 
(iii) Details of the QND calculations using VASP.\\
(iv) The short-ranged ionic structure across LPTs.\\ 
(v) Test of the non-linear Hulth\'en potential.\\
(vi) The compressibility ratio, electrical and thermal conductivities of $l$-Si.\\

\subsection{Brief description of the NPA calculation}
We do not consider well-defined ion configurations with ion-positions
$R_1,R_2\cdots,R_N$ in an $N$-ion simulation cell. Instead, the field ions
 are replaced by their smooth one-body  distribution $\rho(r)$ since  density
 functional theory (DFT) posits that the
 free energy is a functional of just the  one-body density. Thus
 the $N$-body (multi-ion) problem is
replaced by  a  one-body (i.e., one-ion) problem in the sense of DFT. This
becomes exact if the ion correlation functional is exact. In practice
it is  approximate due to uncertainties  in the XC-functionals. The electron
density is also a smooth spherically symmetric distribution centered on
the Si nucleus at the origin. We consider a sphere of radius $R_c\simeq 10 r_{ws}$,
i.e., a ``correlation sphere"
such that all correlation effects have died of an 
the densities $\rho(r), n(r)$ have reduced to their ``bulk'' average values
$\bar{\rho}, \bar{n}$ respectively. The correlation sphere
 reduces to the Debye sphere in the classical
low-density high-$T$ limit. The central Si ion with its associated
electron and ion distributions constitute a neutral object (the NPA) with bound and
free electron states. The NPA can be applied with equal validity from
 the extreme quantum limit of high degeneracy at $T=0$ to the non-degenerate
 classical limit, for fluids or solid phases~\cite{Pe-Be93}. 

Two coupled DFT variational equations appear in the theory of the NPA, since the
total free energy $F([n],\rho]$ is a functional of two densities,
 viz, $n(r)$ and $\rho(r)$~\cite{DWP82}.
\begin{equation}
\label{2dft.eqn}
\frac{\delta F}{\delta n}=0,\;\; \frac{\delta F}{\delta \rho}=0.
\end{equation} 
The first of this equation can be analyzed and re-written as a Kohn-Sham
equation for  electrons subject to a one-body
potential $V_e(r)$ inside the correlation sphere. The second
equation can be treated using classical mechanics and reduces to a
Boltzmann-like equation where the ions are subject to a one-body
potential $V_I(r)$ that includes the nuclear potential, the Poisson
potential and an ion XC-potential $V_{ii}^{xc}(r)$. There are
small e-i XC-potentials that are neglected in this
 treatment~\cite{Furutani90,ilciacco93}. 

The Kohn-Sham calculation yields $n(r)$, as well as the Kohn-Sham eigenstates,
eigen-energies and phase shifts of the ``free-electron'' states. 
 As only a single ion center is used in  NPA calculation, we  do  not have a highly
 inhomogeneous  $N$-center electron distribution
 as in standard DFT simulations (e.g., using  VASP~\cite{VASP}) where several
 hundred ions define a very complex electron distribution in  a simulation cell. 
 Instead, just a single-center spherical electron distribution $n(r)$ and an ion
 distribution $\rho(r)$ dependent
on electron-electron and ion-ion correlation functionals occur in the NPA.
 This much smoother single-center electron density $n(r)$ can be conveniently
 treated using the local density approximation (LDA), as has been found from
  previous work. The electron
exchange correlation functional used~\cite{PDWXC} in the LDA is a
 finite-$T$ functional. Tests using other available functionals, e.g., 
the functionals by Karasiev et al~\cite{KSDT14}, or by Groth et al~\cite{GDS17}
 within the NPA code give essentially indistinguishable results for the range of densities
and temperatures used in this study.

The XC-functional that reduces the multi-ion problem to a single-ion problem
 is known to be highly non-local and hence an LDA approximation is useless. 
However, as ions
are classical particles, several simplifications arise. Although we use the generic name 
XC-potential, the exchange part in taken as zero and there is on a correlation
 functional to deal with. Furthermore, classical statistical mechanics and diagram
 techniques can be used to provide a numerically very convenient and accurate form
 for the ion-correlation functional $F_{ii}^{xc}[\rho(r)]$ if the irreducible
 ``bridge'' diagrams are neglected. When bridge diagrams are important (e.g., in high-density
liquid Al near its melting point where hard-sphere-like packing effects are important),
 they can be included via the Lado-Foils-Ashcroft (LFA)
approach~\cite{LFA83}. However, we have shown ion previous work~\cite{cdwSi20,cdw-carb22}
 that bridge contributions are completely negligible in WDM systems like C,Si where packing
 effects are not important. This because the ion-ion short-ranged structure is determined by
electronic processes near the Fermi surface, and the associated long-ranged
 Friedel oscillations in the ion-ion pair potentials.
The functional derivative of the XC-free energy with respect to
the density is the corresponding ion-XC potential $V_{ii}^{xc}[\rho(r)]$. The latter
 is needed in the density functional equations. It can be given in the
HNC-approximation as
\begin{eqnarray} 
\label{ionVxc.eqn}
\beta V_{ii}^{xc}(r) &=& -\bar{\rho}\int d\vec{r'}\left[h(\vec{r'})+\beta V_I(\vec{r'})\right]
h(|\vec{r}-\vec{r'}|)\\
\beta V_I(r)&=&-\ln[g(r)], \;\; \beta=1/T.
\end{eqnarray}

This $V_{ii}^{xc}(r)$ is a highly non-local quantity, dependent on
$h(r) = g(r)-1$. In solving the coupled equations A1, the
$g_{ii}(r)$ is initially approximated by a cavity $g(r)$ defined
only by $r_{ws}$ ; this $r_{ws}$ is adjusted to self consistency via
the two coupled DFT equations. Then the ion-ion pair-
potential $V_{ii}(r)$, Eq. A6 is calculated via the pseudopotential,
Eq. A3, generated from the final charge density
$n(r)$. Then an HNC equation, or a modified HNC equation with
bridge corrections~\cite{LFA83} is solved to get a very
accurate $g(r)$ instead of the initial cavity-$g(r)$ where only
the cavity radius $r_{ws}$ had been adjusted to be consistent
with the final $\bar{Z}$ that satisfies the Friedel sumrule. This in
effect provides a computationally convenient partial decoupling
of the Kohn-Sham equation for electrons, and
the corresponding classical DFT equation for ions~\cite{eos95}.

The Kohn-Sham calculation provides bound states $\phi_{nl}(r)$ with energies $\epsilon_{nl}$
and continuum states $\phi_{kl}(r)$ with energies $\epsilon_{kl}=k^2/2$. These delocalized
 $\phi_{kl}(r)$ states behave asymptotically like  plane waves with
phase shifts $\delta_{kl}$. They obey the finite-$T$ Friedel sum rule~\cite{DWP82},
which determines the number of free electrons associated with each ion, namely $\bar{Z}$. 
So, at self-consistency, $\bar{Z}=\bar{n}/\bar{rho}$ satisfies the Friedel sumrule, charge
neutrality while being a key quantity in the thermodynamic equilibrium of the fluid,
as it was evaluated via a minimization of
the total free energy via Eq. A1. It has a clear physical meaning~\cite{DWP82} and can
be measured experimentally using Langmuir probes
(for suitable plasma conditions). Indirect methods of determining $\bar{Z}$ are needed
for WDM conditions. 

The Kohn-Sham states of a Si-atom immersed in its ionic and electronic environment
at $\rho$ = 4.6 g/cm$^3$ and $T=0.46$ eV are given in Table~\ref{Si-Enl.tab}.
These results, as well as the Friedel sum evaluated
from the Phase shifts unambiguously show that $\bar{Z}=4$.

\begin{table}
\caption{\label{Si-Enl.tab}The electronic structure at a silicon
nucleus in its fluid state for two cases. Case 1, $\bar{\rho}=$ 4.6 g/cm$^3$ and
$T$= 5300K($\simeq0.46$ eV). Case 2, $\bar{\rho}=$ 4.35 g/cm$^3$ and $T$=1.2 eV. 
The energies are in Rydberg, while the mean orbital radius $<r_{nl}>$ is in 
atomic units. The Wigner-Seitz radii are is 2.537 a.u., and 2.585 a.u. for the two
cases. Hence all bound states are seen to be compactly contained inside
the Wigner-Seitz sphere. The Fermi momenta $k_F$ and $t=T/E_F$ for the two cases are:
 1.2007, 1.1786, and  $t$=0.0235, 0.0635 respectively.
}
\begin{ruledtabular}
\begin{tabular}{lccccc}
case &$n,l$   & $\epsilon_{nl} $ & $<r_{nl}>$  & occupation   \\
\hline\\
1    & 1,0    &  -128.5          & 0.1120      & 1.0$\times$2  \\
2    & 1,0    &  -128.6          & 0.1120      & 1.0$\times$2  \\     
1    & 2,0    &  -8.315          & 0.5738      & 1.0$\times$2  \\
2    & 2,0    &  -8.381          & 0.5733      & 1.0$\times$2  \\
1    & 2,1    &  -5.192          & 0.5416      & 1.0$\times$6  \\ 
2    & 2.1    &  -5.260          & 0.5407      & 1.0$\times$6              
\end{tabular}
\end{ruledtabular}
\end{table}

\subsection{Pseudopotential, Structure data and the first peak of $S(k)$}
The Kohn-Sham-Mermin calculation described above, for a single ion of silicon immersed
in its appropriate environment provides as with the bound density $n_b(r)$ from the
bound wavefunctions, while the continuum wavefunctions provide a density $n_f(r)$ of
free electrons, with $n_f(r)\to\bar{n}$ as $r\to R_c$. Hence the free-electron
density pile up $\Delta n_f(r)=n_f(r)-\bar{n}$.  We take this all-order
result obtained from the Kohn-Sham calculation to be a result of a linear response
to a weak electron-ion pseudo-potential $U_{ei}(k)$,  Then, if $\chi(k)$ is the full interacting
electron response function at the density $]\bar{n}$ and temperature $T$, we have:
\begin{equation}
U_{ei}(k)=\Delta n_f(k)/\chi(k).
\end{equation}
Note that if a point ion of charge $\bar{Z}$ were considered, then its potential
would be 
\begin{equation}
U_{ei}^0(k)=-\bar{Z}V_k, \; V_k=4\pi/k^2.
\end{equation}.
This point-ion interaction is a strong potential and cannot be used in linear-response
theory. The weak pseudopotential that we use has a form factor $W_k$ such that
\begin{equation}
W_k=U_{ei}(k)/U_{ei}^0(k).
\end{equation}
Since the finite-$T$ response function of the uniform electron gas at finite-$T$ inclusive its
local-field correction can be constructed using the electron-electron XC-functional,
$\chi(k)$ is a known quantity. Hence, the Kohn-Sham calculation can be processed to
yield a weak electron-ion pseudopotential. The full details of the calculation may be found
in Refs.~\cite{Pe-Be93, eos95}. Furthermore, given the pseudopotential $U_{ei}(k)$, the
corresponding pair-potential $V_{ii}(k)$ becomes
\begin{equation}
V_{ii}(k)=V_k\bar{Z}^2+|U_{ei}(k)|^2\chi(k), \; V_k=4\pi/k^2.
\end{equation}
This pair-potential can now be used in MD or in the hyper-netted-chain (HNC) equation to
generate the structure factor ion-ion $S(k)$ and the corresponding PDF $g(r)$.
In Ref.~\cite{DW-Yuk22}  it was shown that these pair potentials can be accurately fitted
to a Yukawa-Friedel-Tail (YFT) form of the potential, viz:
\begin{eqnarray}
\label{potform.eqn}
V_{\rm yft} (r)&=&V_{\rm y}(r)+V_{\rm ft}(r),\\
V_{\rm y}&=&(a_{\rm y}/r)\exp(-k_{\rm y}r),\\
V_{\rm ft}&=&(a_{\rm ft}/r^3)\exp(-k_{\rm ft}r)\cos(q_{\rm ft}r+\phi_{\rm ft}).
\end{eqnarray}
The Si-Si pair-potential taken in the form $V_{ii}(x)/T,x=r/r_{ws}$
at the density of 4.35 g/cm$^3$ ($r_{ws}=2.5849$) and 1 eV can be fitted with
the following values of the parameters.
$a_{\rm y}=1076.13, k_{\rm y}=4.18038, a_{\rm ft}=7.60699, k_{\rm ft}=1.18818,
q_{\rm ft}=5.75909, \phi_{\rm ft}= 0.291775.$

\begin{figure}[t]                    
\includegraphics[width=0.96\columnwidth]{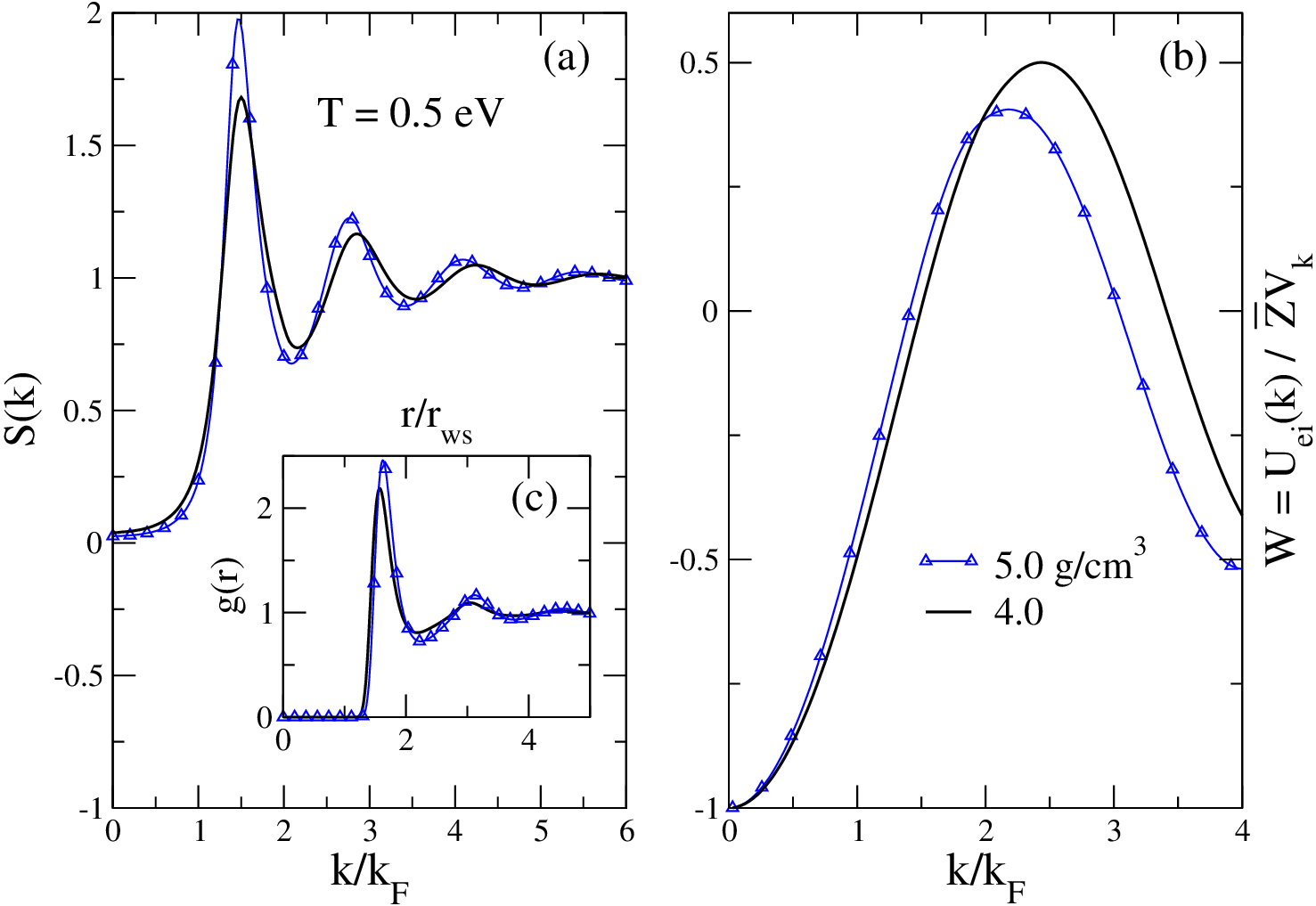}
 \caption{\label{skgrVq.fig}(online color) Panel (a) of the figure shows the Si-Si
$S(k)$ for $\bar{\rho}$=4.0 g/cm$^3$ and $\bar{\rho}$=5 g/cm$^3$ at $T$=0.5 eV. The
position of the first peak of $S(k)$ in $k$-space ($k_1$) remains unchanged in spite
 of the variation in density by 25\%, although the position of the first peak in
 $r$-space in $g(r)$ is affected. Panel  (b) shows that the $k_1$ corresponds
nearly to where the pair potential approximates to a strong point-ion interaction. 
}
\end{figure}

In Fig.~\ref{skgrVq.fig} we display the pseudopotential $U_{ei}(k)$,
 the structure factor $S(k)$
and $g(r)$ for liquid silicon at $T=0.5$ eV, and at two densities,
 viz., 4.0 g/cm$^3$ and
$5.0$ g/cm$^3$. These the XRD and XRTS data of the experiment falls
 within these two densities.
What is note-worthy is that the main peak of $S(k)$ has remained
 ``pinned" at $k_1\simeq 1.499$
which is in fact the value of $k$ which makes the form factor $W_k$
 to be essentially zero.
That is, the first peak of $S(k)$ tries to remain pinned to the
 value of $k$ such that the pair-potential $V_{ii}(k)$ is strongest. 

The relative stability of the short-ranged ionic structure
over this density range is seen also in the g(r) shown
in the inset to Fig. 6. we find that the short-ranged
structure of the ions remains essentially unchanged over
large changes of temperature and density, and even across
LPTs.

\subsection{Details of QMD calculations using VASP}
\label{vasp.sec}
The structure data (i.e., $g(r), S(k)$)obtained from the NPA have been
 established to be in good
 agreement in previous published work, for a variety of materials that include
  Al, Be, C, H, Li,
 Na, Si etc. 
In addition, in the present study also we have checked that our NPA calculations
 agree with results from QMD calculations for test cases. 
For instance, in Fig.~\ref{LPT4p3-1ev.fig} we display the results
obtained for $g(r)$ for $l$-Si at 4.28-4.6 g/cm$^3$ at 1 eV using
the Vienna Ab-initio simulation Package (VASP)~\cite{VASP}.

The VASP simulations are for 64 atoms and carried out
in a standard way~\cite{Vasp-lqdSi}, showing good agreement with the
NPA results. An energy cutoff of $\sim$ 300 eV was employed.
The (Monkhorst Pack) $k$-grid was used and calculations
were done at the $\Gamma$ point (0,0,0). Gaussian smearing
(ISMEAR=0) with a smearing of 0.1 eV was imposed.
The temperature was set using a Nos\'e-Hoover thermostat.
The Purdew-Burke-Ernzerhof XC-functional~\cite{PBE96} was used for
electrons. It was ensured that the occupation in the highest
 energy state is less
than 0.0001 even at the highest temperature studied. 

In a previous QMD study~\cite{Remsing17} of $l$-Si and its PDFs, Remsing et al
studied the convergence of PDFs and other selected properties
for $N=64$, 216, and 512, for the LDA, PBE and SCAN
functionals. Best quantitative accuracy required
 the use of the SCAN functional and N=216, especially for obtaining
good agreement with the melt line near 2.33 g/cm$^3$.  Here
we are considering $l$-Si at higher density and a higher
temperature of 1 eV, favouring the use of smaller simulations.
 The $N=64$ results, though somewhat noisy, were sufficiently well converged
for our purpose of comparisons with the NPA. Increasing $N$ of
course gave better convergence to the NPA values. The larger
$N$ values, as well as the SCAN functional were needed in 
previous studies to capture the correlated phase of the liquid.
In the present study (at the higher $\bar{\rho},T$), the $N=64$
calculations successfully recovered the correlated phase reported
in Fig.~\ref{LPT4p3-1ev.fig}(a) and
(d). Due to our limited computational resources, only
the $N=64$-VASP simulations using  the PBE functional
are presented in this study, although some limited calculations
weere done at $N=125$.

\subsection{The short-ranged ionic
structure across LPTs}
\begin{figure}[t]                    
\includegraphics[width=0.96\columnwidth]{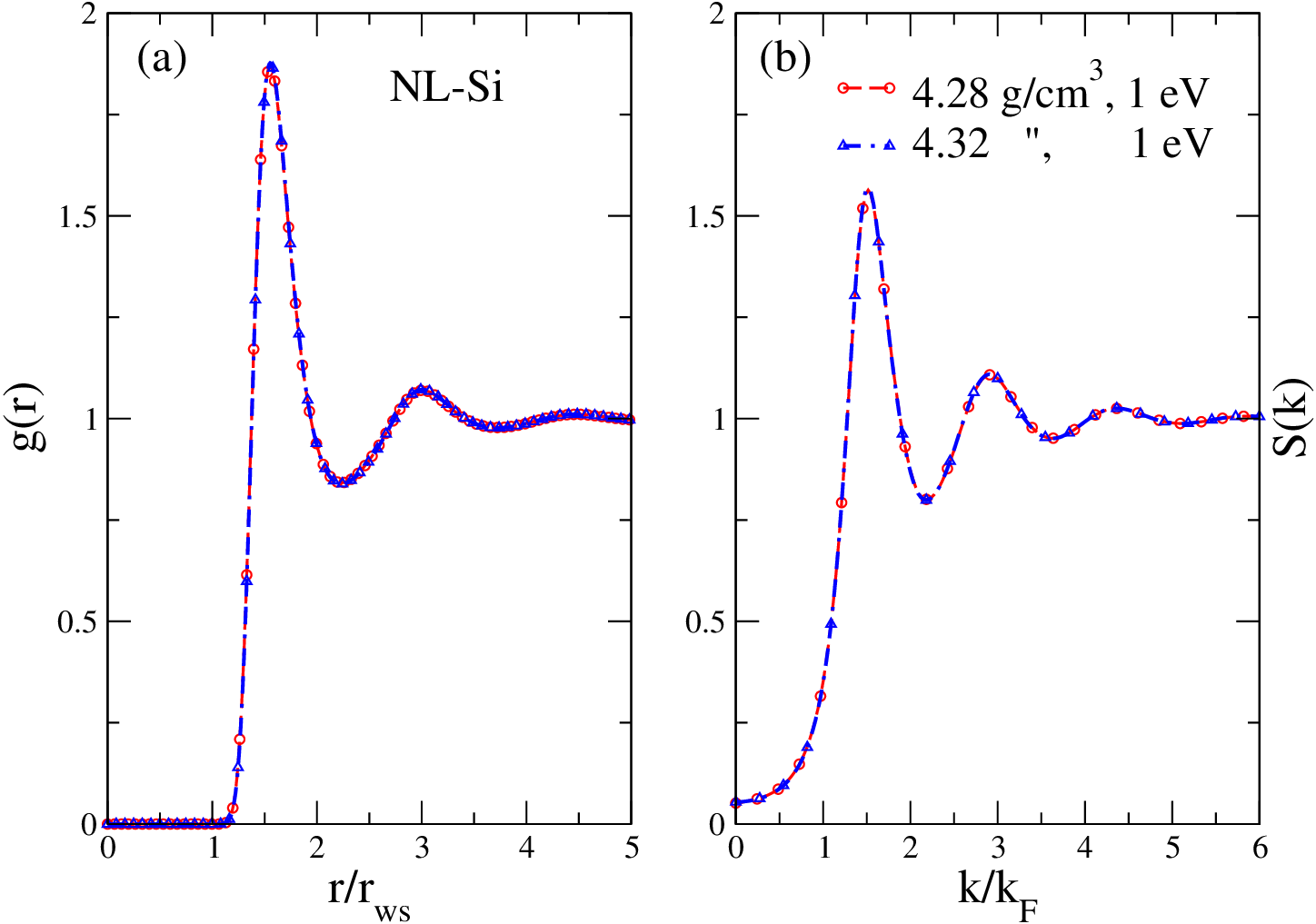}
 \caption{\label{lpt4p3GrSk.fig}(online color) Panel (a) The Si-Si
$g(r))$ for $\bar{\rho}$=4.28 g/cm$^3$ and  4.32 g/cm$^3$ at $T$=1 eV,
calculated using the NPA for the normal liquid phase.
(b) The corresponding $S(k)$ are displayed.
}
\end{figure}

We noted in Sec. A 2 that it is energetically optimal
to preserve the short-ranged ionic structure to position
the electron-ion interactions at a zero of the pseudopotential.
So, structural changes in the normal liquid occur in the
long-ranged tails of the distribution functions, somewhat
as in Martensitic transitions. These transitions appear as
discontinuities in the compressibility and the pressure,
while the short-ranged structure ($r/r_{ws} < 5$) remains
more or less intact. This is demonstrated in Fig.~\ref{lpt4p3GrSk.fig}
We see that the $g(r)$ and $S(k)$ remain unchanged
when going from $\bar{\rho}$ = 4.28 g/cm$^3$ to 4.32 g/cm$^3$
although the pressure and compressibility data show a
discontinuity at  4.3 g/cm$^3$. Similar results are found
for the other LPTs studied here. The LPTs seem to be similar
 to the well-known transition in
l-Si at 2.5g/cm$^3$ near the melting point. However, we
must note that the correlated liquid (CL) phase has so far
only been observed in
numerical simulations, and not in experiments.

This insensitivity of the short-ranged structure to the
LPTs makes them harder to detect via standard simulation methods.
This is probably one reason why the much sought-after LPT in
liquid carbon remained undetected until it
was revealed via detailed NPA calculations~\cite{cdw-carb22}.

\subsection{Tests of the applicability of the Hulth\'en potential}
The analysis of XRD and XRTS data given in Ref.~\cite{Poole24} used the NLH potential
as it is very simple to use, and has met with astrophysical applications in the
past~\cite{Varshni90}. The first-principles potentials for silicon appropriate for
a liquid-environment derived via the NPA can be used to test if the NLH potential
is a satisfactory and simple substitute for use. Fig.~\ref{skgrvr.fig} showed
that NLH differed strongly from the DFT-based potentials calculated at 4.6 g/cm$^3$ and
$T=0.46$ eV. 

As a further test, we used the QMD-generated $g(r)$ for $l$-silicon available for 2.5 g/cm$^3$,
at $T$=1200K, and at $T=1800$K where the SCAN functional has been used in the
 calculations~\cite{Remsing17, cdwSi20}. The QMD generated $g(r)$ agreed very
closely with that from the NPA.  We tried the NLH potential varying its $\bar{Z}$ to
see if the QMD-$g(r)$ could be recovered, but this was not possible. 
This failure of the NLH-potential is a further reason justifying the present study.

\subsection{The compressibility ratio, electrical and thermal conductivites of $l$-Si}
The isothermal compressibility ratio $\xi/\xi_0$ and the electrical conductivity $\sigma$
were displayed in Fig.~\ref{s0.fig} of the main text. Using $\rho$ for
the mean density $\bar{\rho}$ for simplicity, we use the equations given below.
\begin{eqnarray}
\xi&=&\frac{1}{\rho}\left[\frac{\partial \rho}{\partial p}\right]_T=\xi_0\,S(0) \\
\xi_0&=&1/(\rho T),\; \mbox{ideal gas}.
\end{eqnarray}
In Table~\ref{xisigkap.tab} we give a tabulation of
$\xi/\xi_0, \sigma$ together with the corresponding thermal conductivity $\kappa$ calculated
via the Lorentz number, using the approach given in Ref.~\cite{Ln.24}.
As $\bar{Z}$= 4 for silicon, the Spitzer-Harm~\cite{SpHm53}
analysis shows that e-e interactions can be neglected for
 high-$\bar{Z}$ values in calculating transport coefficients.

\begin{table}
\caption{\label{xisigkap.tab} The isothermal compressibility ratio $\xi/\xi_0$ obtained from
the $k\to 0$ limit of the ion-structure factor, the electrical conductivity $\sigma$ calculated
via the T-matrix~\cite{Evens92, Thermophys99}, and the thermal conductivity calculated
via the Lorentz number~\cite{Ln.24} are tabulated below.
}
\begin{ruledtabular}
\begin{tabular}{lllllll}
$T\to$ &   &0.25 eV&  &  &1 eV &  \\
\hline\\
$\bar{\rho}$ & $\xi/\xi_0$ & $\sigma\times10^7$& $\kappa \times10^2$& $\xi/\xi_0$ & $\sigma\times10^7$ & $\kappa\times10^3$\\
 g/cm$^3$ & $\times$0.1 &[Sm] & [Wm] & $\times$0.1 &[Sm] & [Wm] \\
\hline\\
4.20 &  0.1999 & 0.1064	&0.7539 & 0.5390  &  0.1073 &  0.3043 \\
4.22 &  0.1953 & 0.1063	&0.7534 & 0.5338  &  0.1073 &  0.3044 \\         
4.24 &  0.1913 & 0.1062	&0.7529 & 0.5278  &  0.1074 &  0.3045 \\
4.26 &  0.1879 & 0.1061	&0.7523 & 0.5226  &  0.1074 &  0.3047 \\
4.28 &  0.2265 & 0.1059	&0.7508 & 0.5168  &  0.1075 &  0.3048 \\
4.30 &  0.2198 & 0.1058	&0.7503 & 0.5126  &  0.1075 &  0.3049 \\
4.32 &  0.2141 & 0.1057	&0.7496 & 0.5454  &  0.1075 &  0.3050 \\
4.34 &  0.2079 & 0.1056	&0.7490 & 0.5404  &  0.1076 &  0.3050 \\
4.36 &  0.2023 & 0.1056	&0.7483 & 0.5351  &  0.1076 &  0.3051 \\
4.38 &  0.1990 & 0.1055	&0.7476 & 0.5302  &  0.1076 &  0.3052 \\
4.40 &  0.1938 & 0.1054	&0.7469 & 0.5249  &  0.1076 &  0.3052 \\
4.42 &  0.1892 & 0.1053	&0.7462 & 0.5199  &  0.1076 &  0.3053 \\
4.43 &  0.1874 & 0.1052	&0.7458 & 0.5175  &  0.1077  & 0.3053 \\
4.44 &  0.1846 & 0.1052	&0.7458 & 0.5140  &  0.1077  & 0.3053 \\
4.46 &  0.1823 & 0.1050	&0.7446 & 0.5101  &  0.1077  & 0.3053 \\
4.48 &  0.1785 & 0.1049	&0.7438 & 0.5049  &  0.1077  & 0.3053 \\
4.50 &  0.1752 & 0.1048	&0.7430 & 0.4989  &  0.1077  & 0.3053 \\
4.52 &  0.1734 & 0.1047	&0.7421 & 0.4946  &  0.1077  & 0.3053  \\
4.54 &  0.1706 & 0.1046	&0.7413 & 0.4889  &  0.1077  & 0.3053  \\
4.46 &  0.1683 & 0.1045	&0.7404 & 0.4835  &  0.1077  & 0.3053  \\
4.48 &  0.1855 & 0.1044	&0.7400 & 0.4782  &  0.1077  & 0.3053  \\
4.60 &  0.1863 & 0.1043	&0.7391 & 0.4744  &  0.1077  & 0.3053  \\
4.62 &  0.1873 & 0.1041	&0.7382 & 0.4692  &  0.1076  & 0.3052 \\
4.64 &  0.1879 & 0.1034 &0.7372 & 0.4732  &  0.1076  & 0.3052 \\
4.66 &  0.1881 & 0.1038 &0.7362 & 0.4705  &  0.1076  & 0.3051 \\
4.68 &  0.1882 & 0.1037 &0.7352 & 0.4694  &  0.1076  & 0.3051 \\
4.70 &  0.1883 & 0.1036 &0.7342 & 0.4666  &  0.1076  & 0.3059 
\end{tabular}                                
\end{ruledtabular}                           
\end{table}

What is note worthy here is that while the density changes from $\sim$ 4 g/cm$^3$ to
$\sim 5$ g/cm$^3$, i.e., a change of 25\%, the electrical conductivity changes less than
3\% at $T=0.25$ eV, while at 1 eV, the change is less than 0.3\%. The conductivity $\sigma$
has been calculated using the phase-shifts of the continuum states via a T-matrix. However,
we can understand the results more easily via the Ziman formula that uses a pseudopotential.
 The integrand $F(k)$ in the Ziman formula for the conductivity involves a product of $S(k)$
 and the square of  the pseudopotential $|U_{ei}(k)|^2$ in the neighbourhood of $k=2k_F$. 
When the density
changes, as seen from Fig.~\ref{skgrVq.fig}, $F(k)$ remains nearly invariant, with the
changing density accommodated by small but collective changes in distant coordination shells.
This happens in a cooperative manner leading to LPTs, as witnessed by the discontinuities
in the pressure and the compressibility ratio.


\newpage

\begin{thebibliography}{99}
\bibitem{Poole24}
H Poole, M. K. Ginnane, M. Millot, H. M. Bellenbaum, 
G. W. Collins, S. X. Hu,  D. Polsin, R. Saha, J. Topp-Mugglestone,
T. G. White, D. A. Chapman, J. R. Rygg, S. P. Regan,
and G. Gregori.
Physical Review Research {\bf 6}, 023144 (2024)
DOI: 10.1103/PhysRevResearch.6.023144

\bibitem{Ng05}
 A. Ng, T. Ao, F. Perrot, M.W.C. Dharma-wardana, M.E. Foord,
Laser and particle beams, {\bf 23}, 527-537 (2005).

\bibitem{McBride-Si-19}
E. E. McBride, A. Krygier, A. Ehnes, E. Galtier, M. Harmand, Z. Kon\^{o}pkov\'{a},
{\it et al., }
Nature Phys. {\bf 15}, 89-94 (2019). 


\bibitem{Si-Aptekar79}
L. I. Aptekar, Sov. Phys. Dokl. {\bf 24}, 993 (1979).

\bibitem{Stich89}
I. \v{S}tich, R. Car, and M. Parrinello, 
Phys. Rev. Lett. {\bf 63}, 2240 (1989).


\bibitem{DWP-carb90} M. W. C. Dharma-wardana and F. Perrot, Phys. Rev. Lett.,
 {\bf 65}, 76 (1990).

\bibitem{OkadaSi12}
J. T. Okada, P. H.-L. Sit, Y. Watanabe,  Y. J. Wang,  B. Barbiellini,  T. Ishikawa,
{\it et al., }
Phys. Rev. Lett. {\bf 108},  067402  (2012).


\bibitem{CPP-carb18}
M. W. C. Dharma-wardana,
Contrib. Plasma Phys. {\bf 58} 128-142 (2018).



\bibitem{sxhu2017}
S. X. Hu, R. Gao, Y. Ding, L. A. Collins, and J. D. Kress,
 Phys. Rev E {\bf 95}, 043210 (2017).

\bibitem{cdwSi20}
M.W.C. Dharma-wardana, Dennis D. Klug, and Richard C. Remsing
Phys. Rev. Lett. {\bf 125}, 075702 (2020). 
doi: 10.1103/PhysRevLett.125.075702


\bibitem{SastryAngel03}
S. Sastry and C. A. Angell, Nat. Mater. {\bf 2}, 739 (2003).


\bibitem{Ashwin04}
S. S. Ashwin, U. V. Waghmare and  Srikanth Sastry,
Phys. Rev. Lett. {\bf 92}, 175701 (2004).

\bibitem{Morishita04}
T. Morishita, Phys. Rev. Lett. {\bf 93}, 055503 (2004)


\bibitem{McMillan05}
P. F. McMillan, M. Wilson, D. Daisenberger and  D. Machon,
Nat. Mater. {\bf 4}, 680  (2005).

\bibitem{Morishita06}
T. Morishita, Phys. Rev. Lett. {\bf 97}, 165502 (2006).

\bibitem{Beaucage05}
 P. Beaucage and N. Mousseau, J. Phys. Condens. Matter,
{\bf 17}, 2269 (2005).

\bibitem{Daisen07SiPhDia}
Dominik Daisenberger, Mark Wilson, Paul F. McMillan, R. QuesadaCabrera,
Martin C. Wilding, Denis Machon,
Phys. Rev. B  {\bf 75}, 224118 (2007).

\bibitem{GaneshSiLPPT-09}
 P. Ganesh and M. Widom, Phys. Rev. Lett. {\bf 102}, 075701
(2009).

\bibitem{Baye10}
M. Baye,
Martin Beye, Florian Sorgenfrei, William F. Schlotter, Wilfried Wurth,
 and Alexander F\"{o}hlisch, 
PNAS, {\bf 28}, 16772 (2010).

\bibitem{Vasisht11}
V. V. Vasisht, S. Saw, and S. Sastry, Nat. Phys. {\bf 7}, 
549 (2011).

\bibitem{Remsing17}
Richard C. Remsing,  Michael L. Klein and Jianwei Sun, 
Phys. Rev. B {\bf 96}, 024203 (2017).

\bibitem{Remsing18}
Richard C. Remsing,  Michael L. Klein and Jianwei Sun, 
Phys. Rev. B {\bf 97}, 140103(R) (2018).


\bibitem{ABINIT}
X.Gonze and C. Lee, Computer Phys. Commun. \textbf{180}, 2582-2615 (2009).

\bibitem{VASP}
G. Kresse and J. Furthm\"{u}ller, Phys. Rev. B \textbf{54}, 11169 (1996).



\bibitem{Schorner23}
M. Sch\"{o}rner, M. Bethkenhagen, T. D\"{o}ppner, D. Kraus, L. B.
Fletcher, S. H. Glenzer, and R. Redmer, 
ing spectra from density functional theory molecular dynamics
simulations based on a modified chihara formula, 
Phys. Rev. E {\bf 107}, 065207 (2023).

\bibitem{Witte17}
B. B. L. Witte, M. Shihab, S. H. Glenzer, and R. Redmer,
Phys. Rev. B {\bf 95}, 144105 (2017).



\bibitem{Varshni90}
 Y. P. Varshni, 
Phys. Rev. A {\bf 41}, 4682 (1990).

\bibitem{Vorberger13}
J. Vorberger, D. O. Gericke, and W.-D. Kraeft, 
High Energy Density Phys.{\bf 9}, 448 (2013).

\bibitem{Ebeling20}
W.Ebeling, H. Reinholz, and G. R\"{o}pke, 
 Eur. Phys. J.: Spec. Top. {\bf 229},3403 (2020).

\bibitem{LLvol10}
E. M. Lifshitz and L. P. Pitaevaskii, {\it Physical Kinetics}, Pergamon, New York (1981),
section 44.


\bibitem{DWP82}
M. W. C. Dharma-wardana and F. Perrot, 
Phys. Rev. A {\bf 26}, 2096  (1982).

\bibitem{Pe-Be93} 
 F. Perrot,  Phys. Rev. E {\bf 47}, 570 (1993).

\bibitem{eos95}
F. Perrot and M.W.C. Dharma-wardana,
Phys. Rev. E. {\bf 52}, 5352 (1995). 


\bibitem{DagensNaLi72}
L. Dagens, J. Phys. C: Solid State Physics, {\bf 5}, 2333 (1972).


\bibitem{DW-Yuk22}
M. W. C. Dharma-wardana, Lucas J. Stanek, and Michael S. Murillo
Phys. Rev. E {\bf 106}, 065208 (2022).


\bibitem{Whitley15}
H. D. Whitley, 
D. M. Sanchez , S. Hamel , A. A. Correa,
and L. X. Benedict, Contrib. Plasma Phys. {\bf 55}, 390 (2015).

\bibitem{ercolessi1994interatomic}
F. Ercolessi and J. B. Adams.
EPL (Europhysics Letters), {\bf 26}(8), 583 (1994).

\bibitem{Brommer2015Potfit}
P. Brommer, A. Kiselev, D. Schopf, P. Beck, J.
Roth, and H.-R. Trebin.
Modelling and Simulation in Materials Science and Engineering, 
{\bf 23}(7), 074002, (2015).

\bibitem{Stanek21}
 Lucas J. Stanek, Raymond C. Clay III, M. W. C. Dharma-wardana,
Mitchell A. Wood, Kristian R. C. Beckwith, and Michael S. Murillo,
Phys. Plasmas {\bf 28}, 032706 (2021).



\bibitem{kraus13}
D. Kraus,
J. Vorberger, D. O. Gericke, V. Bagnoud, A. Blazevic, W. Cayzac,
A. Frank, G. Gregori, A. Ortner, A. Otten, F. Roth, G. Schaumann,
D. Schumacher, K. Siegenthaler, F. Wagner, K. Wunsch, and M. Roth,
Phys. Rev. Let.  {\bf 111}, 255501 (2013).


\bibitem{SCAN2013}
J. Sun,  B. Xiao, Y. Fang, {\it et al.}
Phys. Rev. Lett. {\bf 111}, 106401 (2013).

\bibitem{KadauFe2002}
K. Kadau, T. C. Germann, P. S. Lomdahl, and B. L. Holian.
Science, {\bf 296}, 1681 (2002)
Shock waves.




\bibitem{cdw-Utah12}
M. W. C. Dharma-wardana, 
Phys. Rev. E {\bf 86}, 036407 (2012).


\bibitem{xrt-Harb16}
L. Harbour, M. W. C. Dharma-wardana, D. Klug and L. Lewis, 
Physical Review E {\bf 94}, 053211, (2016).

\bibitem{Andreoni76}
W. Andreoni, Pys. Rev. B {\bf 14}, 4247 (1976).

\bibitem{cdw-carb22}
M. W. C. Dharma-wardana and Dennis D. Klug,
 Phys. Plasmas {\bf 29}, 022108  (2022); doi: 10.1063/5.0077343

\bibitem{Furutani90}
 F. Perrot, Y. Furutani and M.W.C. Dharma-wardana,
Phys. Rev. A {\bf 41}, 1096-1104 (1990)

\bibitem{ilciacco93}
E. K. U. Gross, and R. M. Dreizler,
{\it Density Functional Theory},
 NATO ASI series, {\bf 337}, 625
 Plenum Press, New York (1993).


\bibitem{PDWXC}
F. Perrot and M. W. C. Dharma-wardana, Phys. Rev. B {\bf 62}, 16536 (2000);
{\it Erratum: } {\bf 67}, 79901 (2003); arXive-1602.04734.

\bibitem{KSDT14}
Karasiev, V. V.;  Sjostrom, T.;  Dufty, J.;  and  Trickey,
 S. B.; {\it Phys. Rev. Lett.} {\bf 2014}
{\em 112} 076403.

\bibitem{GDS17}
S. Groth, T. Dornheim, T. Sjostrom, F.D. Malone, W. Foulkes, M. Bonitz,
Phys. Rev. Lett. {\bf 119} (13)  135001 (2017).

\bibitem{LFA83}
F. Lado, S. M. Foiles and N. W. Ashcroft,  Phys. Rev. {\bf A 26}, 2374 (1983).


\bibitem{Vasp-lqdSi}
Details for a standard MD simulation of $l$-Si using VASP are given in
\url{https://www.vasp.at/wiki/index.php/Liquid_Si_-_Standard_MD}


\bibitem{Evens92}
R. Evans,  {\it Fundamentals of Inhomogeneous Fluids} (New York: Dekker) p
85 (1992).

\bibitem{Thermophys99}
F. Perrot {\it et al.}
and M. W. C. Dharma-wardana,
Int. J.  of Thermophys, {\bf 20},1299 (1999).

\bibitem{PBE96}
J. P. Perdew, K. Burke, and M. Ernzerhof, Phys. Rev.
Lett. {\bf 77}, 3865 (1996).

\bibitem{Ln.24}
M. W. C. Dharma-wardana (unpublished)
preprint: https://arxiv.org/abs/2404.19692

\bibitem{SpHm53}
L. Spitzer and R. H\"{a}rm, Phys. Rev. {\bf 89}, 977 (1953).


\end{thebibliography}
\end{document}